\documentclass[twoside,twocolumn,9pt]{article}
\pdfoutput=1
\usepackage{extsizes}
\usepackage[super,sort&compress,comma]{natbib} 
\usepackage[version=3]{mhchem}
\usepackage[left=1.5cm, right=1.5cm, top=1.785cm, bottom=2.0cm]{geometry}
\usepackage{balance}
\usepackage{mathptmx}
\usepackage{sectsty}
\usepackage{graphicx} 
\usepackage{lastpage}
\usepackage[format=plain,justification=justified,singlelinecheck=false,font={stretch=1.125,small,sf},labelfont=bf,labelsep=space]{caption}
\usepackage{float}
\usepackage{fancyhdr}
\usepackage{fnpos}
\usepackage[english]{babel}
\addto{\captionsenglish}{}
\usepackage{array}
\usepackage{droidsans}
\usepackage{charter}
\usepackage[T1]{fontenc}
\usepackage[usenames,dvipsnames]{xcolor}
\usepackage{setspace}
\usepackage[compact]{titlesec}
\usepackage{hyperref}

\usepackage{epstopdf}%This line makes .eps figures into .pdf - please comment out if not required.

\begin{document}

%%%TITLE, AUTHORS AND ABSTRACT%%%
\twocolumn[
\begin{@twocolumnfalse}
\noindent\Large{\textbf{$1/f$ noise spectroscopy and noise tailoring of nanoelectronic devices}}\\

\noindent\large{Zolt\'an Balogh,\textit{$^{a,b}$}, Gr\'eta Mezei,\textit{$^{a,b}$} L\'aszl\'o P\'osa,\textit{$^{a,c}$}, Botond  Sánta,\textit{$^{a,b}$}, Andr\'as Magyarkuti,\textit{$^{a,b}$} and Andr\'as Halbritter$^{\ast}$\textit{$^{a,b}$}}\vspace{0.6cm}

\textit{$^{a}$~Department of Physics, Budapest University of Technology and Economics, Budafoki \'ut 8, H-1111 Budapest, Hungary. E-mail: halbritt@mail.bme.hu}\\
\textit{$^{b}$~MTA-BME Condensed
Matter Research Group, Budafoki \'ut 8, H-1111 Budapest, Hungary}\\
\textit{$^{c}$~Institute of Technical
Physics and Materials Science, Centre for Energy Research, Konkoly-Thege M. \'ut 29-33, 1121 Budapest, Hungary}\vspace{0.6cm}

\noindent\normalsize{In this paper, we review the $1/f$-type noise properties of nanoelectronic devices focusing on three demonstrative platforms: resistive switching memories, graphene nanogaps and single-molecule nanowires. The functionality of such ultrasmall devices is confined to an extremely small volume, where bulk considerations on the noise loose their validity: the relative contribution of a fluctuator heavily depends on its distance from the device bottleneck, and the noise characteristics are sensitive to the nanometer-scale device geometry and the details of the mostly non-classical transport mechanism. All these are reflected by a highly system-specific dependence of the noise properties on the active device volume (and the related device resitance), the frequency, or the applied voltage. Accordingly, $1/f$-type noise measurements serve as a rich fingerprint of the relevant transport and noise-generating mechanisms in the studied nanoelectronic systems. Finally, we demonstrate that not only the fundamental understanding and the targeted noise suppression is fueled by the $1/f$-type noise analysis, but novel probabilistic computing hardware platforms heavily seek well tailorable nanoelectric noise sources.}
\end{@twocolumnfalse} \vspace{0.6cm}]

\section{Introduction}
Noise is commonly considered as a disturbing factor. In traditional electrical engineering, it is indeed a challenge to reduce the intrinsic noise of the devices as much as possible.\cite{Balandin1999,Balandin2002}  On the other hand, it is also known that noise studies provide a rich source of information about the devices under study, which cannot be extracted from mean conductance data.\cite{Kogan1996} This was remarkably well demonstrated in the field of mesoscopic physics,  where shot noise measurements\cite{Beenakker2003} resolved non-unity quasiparticle charge values;\cite{Jehl2000,de-Picciotto1997} the crossover between classical and quantum chaos;\cite{Oberholzer2002} the distinction between fermionic and bosonic behavior;\cite{Henny296,Oliver299} and valuable information about the quantum conductance channels of the devices.\cite{Brom1999,Djukic2006,Vardimon2015} In this paper, we treat an other fundamentally important noise-type, the so-called $1/f$-type noise:\cite{Kogan1996,Balandin2013} we review, how $1/f$-type noise studies deliver a deeper understanding of the transport properties and the dominant noise sources in ultrasmall nanoelectronic devices. 

Throughout the paper we briefly review the basics of $1/f$-type noise analysis (Sec. \ref{basics}), we discuss how the analysis of the noise's resistance and frequency scaling (Secs. \ref{resscale}, \ref{freqscale}) as well as nonlinear noise spectroscopy (Sec. \ref{nonlin}) are applied as microscopic tools to gain information about atomic-scale processes in nanoelectronic devices, and finally we demonstrate the merits of \emph{noise tailoring},\cite{Santa2021} i.e. the novel perspective of harvesting device noise as a computational resource in probabilistic computing hardware machines (Sec. \ref{tailor}).\cite{Borders2019,Cai2020}

Our discussions basically rely on three fundamentally different nanoelectronic platforms lying in the focus of our own interest: resistive switching memory devices (or memristors),\cite{Zidan2018,Xia2019,WaserBook} graphene nanogap devices,\cite{Nef2014,JanAMol2015,Heerema2016,QizhiXu2017,Posa2017,Sarwat2017,ElAbbassi2019} and single-molecule nanowires,\cite{Elke_review,Aradhya2013} and mostly we demonstrate the concepts of the analysis by replotting our own data in a simplified, illustrative fashion. We emphasize, however, that these results are all inspired by a broad range of similar studies in the literature, and furthermore, the whole field relies on the solid background of past $1/f$-type noise studies on bulk systems.\cite{Kogan1996} As the central goal of this somewhat tutorial review, we wish to demonstrate that $1/f$-type noise studies become even more interesting as the active volume of the devices enter the ultramsall, nanometer-scale regime, where the noise's scaling with the device resistance, frequency or voltage becomes highly specific to the geometry, transport mechanism and the dominant noise-generating processes of the actual device.

\section{Noise basics}
\label{basics}

\begin{figure}[t]
  \centering{\includegraphics[width=0.48\textwidth]{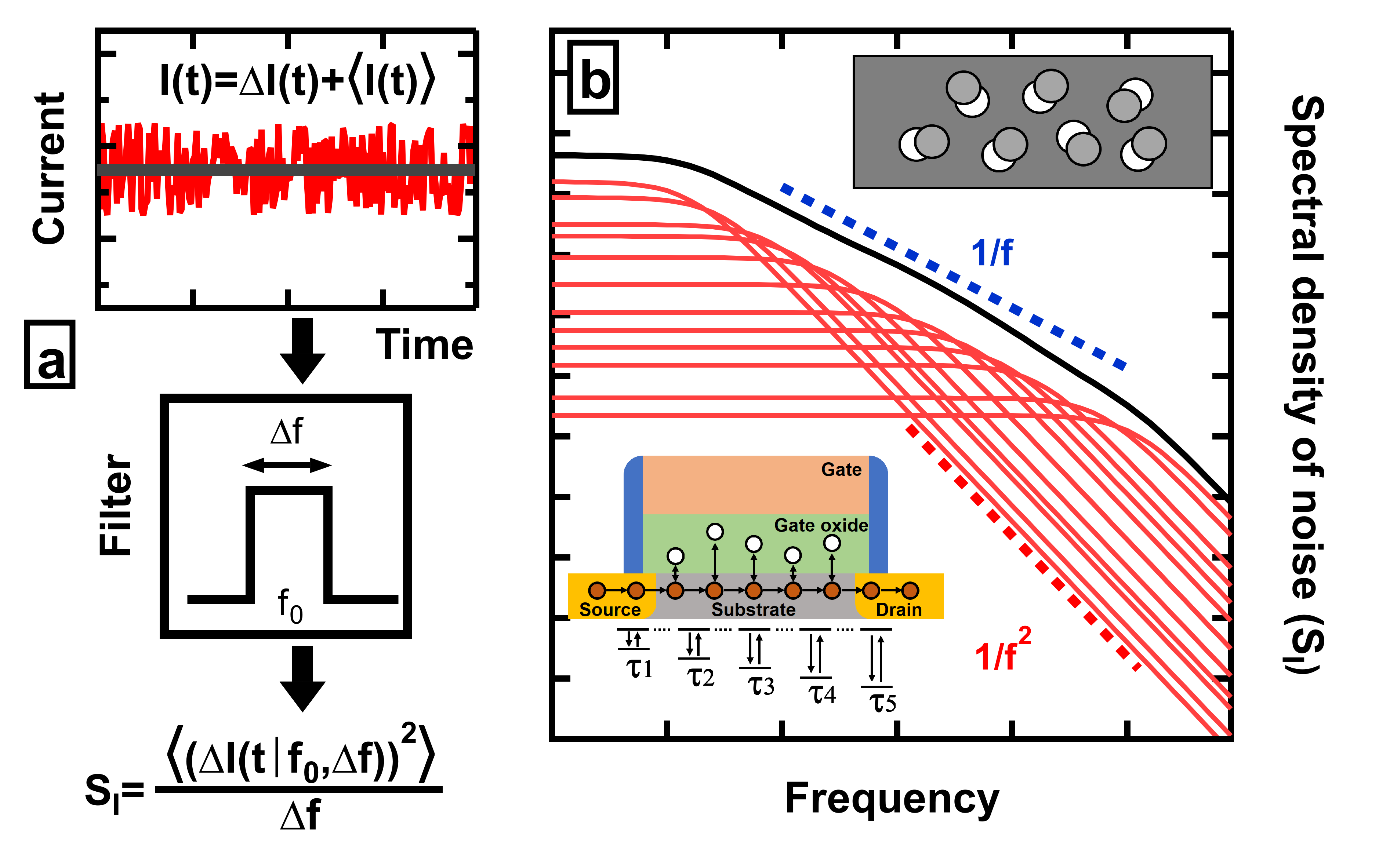}
  \caption{{\bf Noise basics.} (a) A fluctuating $I(t)$ current can be decomposed to its mean value and the $\Delta I(t)$ noise around the mean value. The mean squared deviation of the latter within a $\Delta f$ band around the $f_0$ central frequency normalized to $\Delta f$ gives us the $S_I$ spectral density of the current noise. (b) The $1/f$-type noise of an electronic device is considered as the conducting region's resistance fluctuation. In a semiconductor device, like a field effect transistor this is commonly related to the trapping/de-trapping of electrons in the nearby oxide (see bottom inset). In metals rather the scattering of electrons on dynamical fluctuators, like atoms fluctuating between metastable positions are responsible for the resistance noise (see illustration in the top inset). A single noise-generating item with a characteristic $\tau$ fluctuation time-scale yields a Lorentzian spectrum with constant ($1/f^2$) frequency scaling in the low (high) frequency limit (see red curves). The added contribution of many fluctuators with various time-constants yields a shallower frequency scaling. In this particular figure the spectra of 12 fluctuators are superimposed (black line) such that the $\log(\tau)$ values are uniformly distributed random variables within a predefined region. In the corresponding frequency region the envelope spectrum is close to the $1/f$ frequency scaling (see the blue dotted line). }
  \label{Fig1}}
\end{figure}

Instead of being a \emph{disturbing} signal component (like a noisy neighbor, or a $50\,$Hz pickup), spectrum analysis rather considers noise as a special type of signal with a broad and rather continuous spectrum including contributions from all frequencies. For such a signal, the mean squared deviation within a small $\Delta f$ bandwidth around the central frequency $f_0$ scales with the bandwidth, and the proportionality factor is the \emph{spectral density of noise}, or the  \emph{noise power spectrum density} (PSD). For current noise, this relation reads as $\left<\left(\Delta I(t|f_0,\Delta f)\right)^2\right>=S_I(f_0)\cdot \Delta f$, which is considered as an \emph{experimental definition} of noise\cite{Kogan1996} (see the illustration in Fig.~\ref{Fig1}a). 

From a theoretician's point of view, the description of noise is usually approached through correlation functions. Fortunately, these approches are equivalent: according to the
Wiener–Khinchin theorem, the above defined $S_I(f)$ spectral density is twice the Fourier-transform of the current-current correlation function,\cite{Kogan1996} $C_I(t_2-t_1)=\left<\Delta I(t_1) \cdot \Delta I(t_2)\right>$. From a practical point of view, the noise PSD is usually evaluated from the Fourier transform of the measured current fluctuation as $S_I(f)=\left<(2\Delta t/N)\cdot\left|\sum_{n=0}^{N-1}\Delta I(n\Delta t)\exp(-i2\pi f n \Delta t)\right|^2\right>$, where $\Delta t$ is the time-delay between the adjacent acquisition events, and $N$ is the number of acquired data points. Voltage noise is similarly defined, and in the linear, resistive transport regime its spectral density transforms as $S_V=R^2\cdot S_I$, where $R$ is the resistance.

Thermal fluctuations of a resistor yield a voltage noise with the frequency-independent spectral density of $S_V=4kTR$, where $k$ is Boltzmann's constant, $T$ is the absolute temperature and $R$ is the resistance. This so-called \emph{Johnson-Nyquist noise}\cite{Johnson1928,Nyquist1928} or \emph{thermal noise} has fundamental role in temperature metrology, and it is a key restriction for the base noise level in ultrasensitive measurements. 

Another fundamental noise type is the \emph{shot noise}. If the quantized $e$ electron charges are transmitted through a barrier independently from each other, the Poissonian distribution of the transmission events yields a frequency-independent current noise density which is proportional to the mean current, $S_I=2e\left< I \right>$. This principle was applied by Walter H. Schottky to determine the elementary charge from the current noise of vacuum tubes.\cite{Schottky1918} The same shot noise formula applies to electrons passing through a semiconductor heterojunction at low currents, or through a tunneling barrier. In more complex nanoelectronic circuits, the criterion of independent transmissions is not satisfied, and the amplitude of the shot noise carries valuable information about the quatum conductance channels of the investigated nanostructures.\cite{Brom1999,Djukic2006,Vardimon2015}

In this review, we focus on a third fundamental noise type, the $1/f$-type noise.\cite{Kogan1996,Balandin2013} In electronic devices the $1/f$-type noise is usually considered as a resistance noise which is generated by dynamical fluctuations in or around the device’s active volume (e.g. scattering on dynamical defects inside the conduction channel, occupation/emptying of charge-trap states around the conduction channel, etc., as illustrated in the insets of Fig.~\ref{Fig1}b). A single fluctuator with a characteristic fluctuation time constant, $\tau$ results in a resistance-resistance correlation function decaying with this time constant: $\left< \Delta R(t_1)\cdot \Delta R(t_2)\right>\approx\left <(\Delta R)^2\right>\cdot \exp(-(t_2-t_1)/\tau)$. The spectral density of this resistance noise is obtained from this correlation function by Fourier transformation: $S_R (f)=4\left<(\Delta R)^2 \right>\tau/(1+(2\pi f)^2 \tau^2 )$. This Lorentzian function yields a frequency independent noise in the low-frequency region $(2\pi f\ll 1/\tau)$ and a $1/f^2$ frequency dependence at high frequency $(2\pi f\gg 1/\tau)$. However, if several fluctuators with a broad distribution of time constants are superimposed, the summed noise exhibits a shallower distribution with $S_R (f)\sim1/f^\gamma$, where the $\gamma$ exponent is typically close to 1 (see Fig.~\ref{Fig1}b). 

\begin{figure}[h!]
  \centering{\includegraphics[width=0.48\textwidth]{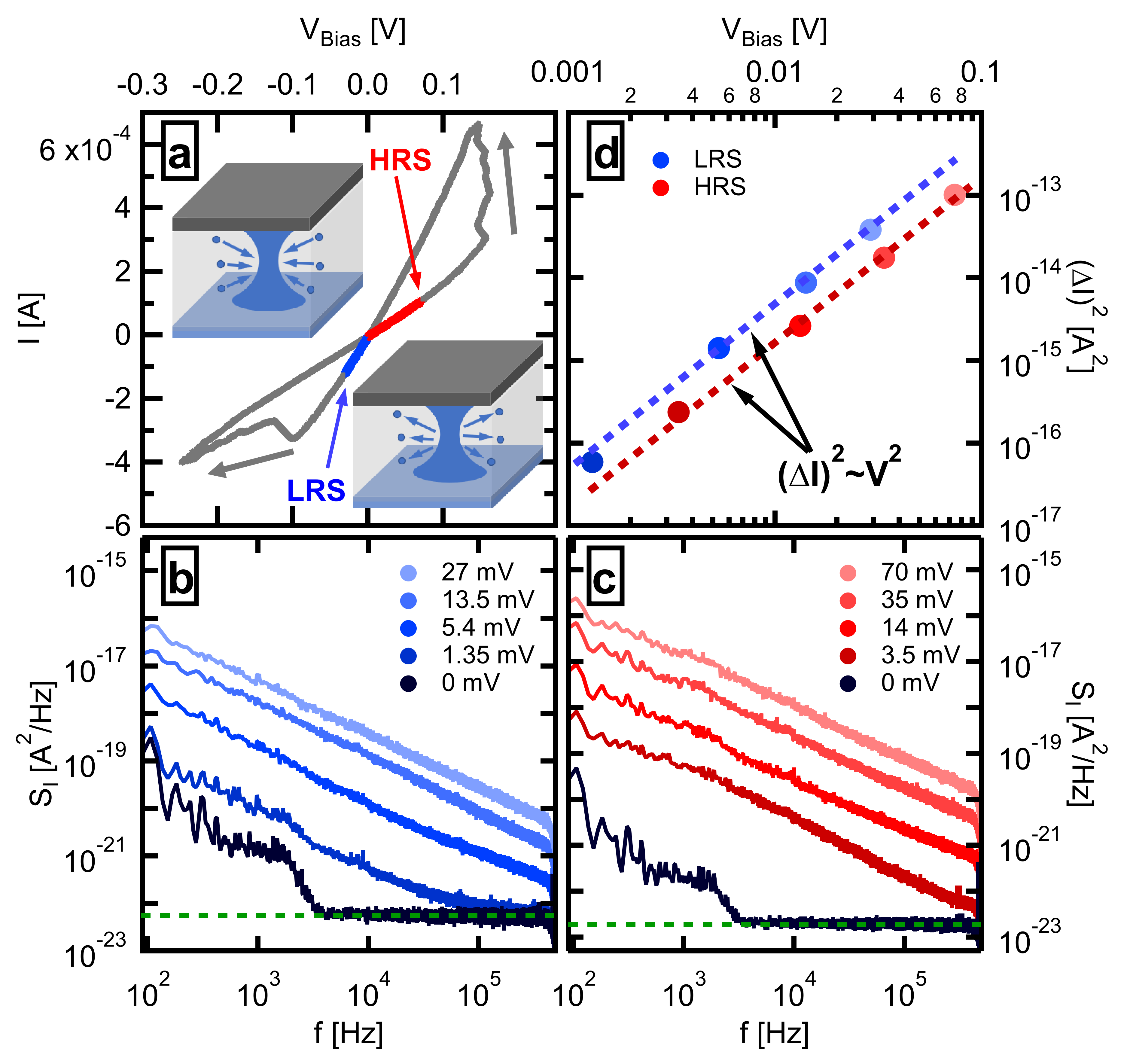}
  \caption{{\bf A typical $1/f$-type noise measurement.} (a) The $I(V)$ characteristic of a silver suflide resitive switching junction. Within the Ag$_2$S matrix a silver conducting filament is formed. Above a certain positive threshold voltage, this filament grows wider due to electrochemical metallization, and thereby the device switches from a high resistance state (HRS) to a low resistance  state (LRS). At opposite polarity, a reversed process occurs, and the device switches back to its HRS. The insets illustrate the widening and shrinking of the filament. The resistance of the both states is analog tunable by the amplitude of the applied triangular driving voltage signal. The further panels show the noise measurement in the linear part of the HRS and LRS (see the red and blue segments respectively). (b,c) The raw noise spectra in the LRS (b, blue spectra) and the HRS (c, red spectra). In both cases, the black curves show the spectrum at zero bias yielding the thermal noise background at high frequency and instrumental background at low frequency. The thermal noise level for the HRS and LRS resistances are shown by green dashed lines in both panels. As the voltage is increased $1/f$-type spectra evolve from the thermal noise background with an average exponent of $\gamma=1.12\pm 0.12$. (d) The noise power integrated for the $100\,$Hz$-500\,$kHz frequency band as a function of the bias voltage for both states (HRS-red, LRS-blue). Before integration the zero bias background noise level is subtracted. The log-log plot exhibits a slope of $2$ for both states, i.e. the expected $V^2$ voltage-scaling of the noise amplitude is satisfied (see text). The data are taken from Ref. \citenum{Santa2019} after converting the voltage noise data measured by cross-correlation technique to current noise data.}
  \label{Fig2}}
\end{figure}

Although we consider $1/f$-type noise as a resistance fluctuation, in experiments, voltage or current noise is more commonly measured, which can be calculated from resistance or conductance noise as: $\left<(\Delta V)^2\right>=\left<(\Delta R)^2\right>\cdot I^2$ or $\left<(\Delta I)^2\right>=\left<(\Delta G)\right>^2\cdot V^2$. Here, the mean squared deviations are considered as the integrated noise powers within the frequency band of the measurement, like $\left<(\Delta I)^2\right>=\int S_I(f)\mathrm{d}f$.  This means that the conventional squared dependence of the noise power on the driving signal amplitude is not related to any current or voltage-induced excitation, it is simply the readout of the steady state resistance/conductance noise according to Ohm’s law. Fig.~\ref{Fig2} exemplifies the typical $V^2$ voltage dependence of the $1/f$-type current noise emerging from the thermal noise background in Ag$_2$S resistive switching memories (see the caption for more details). According to these considerations, a resistive device in the linear transport regime exhibits \emph{voltage-independent} relative current, voltage, conductance or resistance fluctuations, which are equal to each other according to Ohm's law: $\Delta I/I=\Delta V/V=\Delta G/G=\Delta R/R$. Note, that here the fluctuation of a quantity is considered as its standard deviation, like $\Delta I=\sqrt{ \left<(\Delta I)^2\right>}$. 

To translate the above summarized concepts to actual measurements, carefully designed measurement apparatus are required. This includes thorough shielding and grounding, and properly chosen low-noise sourcing and measurement units. In case of voltage noise measurements a cross-correlation measurement scheme can be applied, where the noise is determined from the cross-spectrum of two parallel amplifiers, and thereby the input noise of the amplifiers is cancelled.\cite{Kumar1996} In noise measurements it is essential to acquire the data applying an anti-aliasing filter, and the low-frequency noise spectra are usually calculated by a simple FFT algorithm applying an averaging over the spectra of subsequent time traces.    

After these introductory remarks, we turn to the analysis of the low-frequency noise properties of nanoelectronic devices. In such structures, i.e. where the operation is governed by an extremely small volume, the $1/f$-type noise exhibits a non-obvious, device-specific scaling with the various parameters, like resistance, frequency and voltage. The analysis of these dependencies delivers essential information about the dominant noise-generating microscopic processes, as detailed in the following sections.

\section{Resistance scaling of the noise}
\label{resscale}
In a \emph{bulk sample}, simple considerations can be made on the scaling of the relative noise level with the sample volume.\cite{Wu2008} Assuming $N$ independent dynamical fluctuators homogeneously distributed in the sample volume $(V)$, we can consider an elementary volume of $l^3_F=V/N$ including a single fluctuator on average. The $R_F$ resistance and the $\left<(\Delta R_F)^2\right>$ mean squared resistance fluctuation of this elementary volume adds to an overall relative mean squared resistance fluctuation for the entire sample, $\left<(\Delta R)^2\right>/R^2=(l^3_F/V) \cdot \left<(\Delta R_F)^2\right>/R_F^2$. This result simply relies on the additive nature of the resistances (conductances) and mean squared resistance (conductance) fluctuations for serial (parallel) connected elementary volumes.\cite{Wu2008} 
%Note, that $\left<(\Delta R_F)^2\right>$ is the integrated $S_{R_F}$ spectral density of noise for the frequency band of the measurement. 
If the noise power averaged for all the elementary volumes follows a  truly $1/f$ frequency scaling $\left(\left<S_{R_F}\right>=\alpha_F/f\right)$, the above relation yields $S_R/R^2=(l^3_F/V) \cdot \alpha_F/(f\cdot R_F^2)$ for the resistance noise power of the entire sample, which is closely related to the empirical Hooge's law describing the $1/f$ noise in bulk samples.\cite{Hooge1969,Hooge1981,Kogan1996} These results tell us that the relative noise level scales inversely with the device volume in a bulk sample including volume-distributed fluctuators.

\begin{figure}[h!]
  \centering{\includegraphics[width=0.48\textwidth]{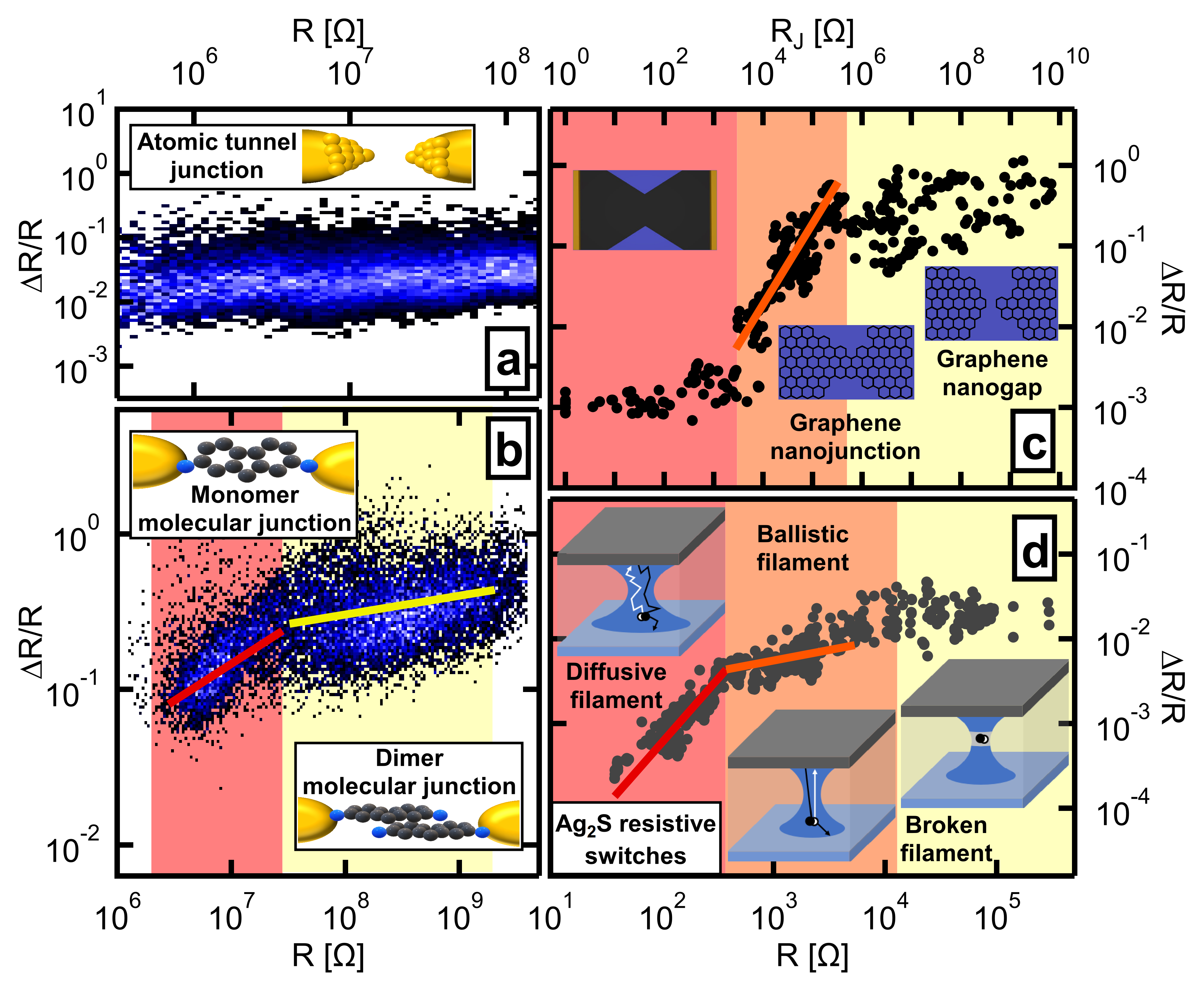}
  \caption{{\bf Resistance scaling of the noise in atomic tunnel junctions, single-molecule junctions, graphene nanogaps and resistive switching memory junctions.} (a) Atomic tunnel junctions exhibit an almost resistance independent relative resistance fluctuation (own measurements, similar to the data in Ref. \citenum{Adak2015}). (b) Gold-2,7-diaminofluorene-gold molecular junctions exhibit two distinct molecular configurations. The low resistance configuration (light red region) is related to a through-bond coupled single-molecule configuration (see top inset). This is confirmed by the $\Delta R /R\sim R^{0.475}$ resistance scaling (see the red fitting line). The high resistance (yellow) regime is related to through-space coupled dimer junctions with a weak resistance dependence of the noise ($\Delta R /R\sim R^{0.12}$, see the yellow fitting line). Data taken from Ref. \citenum{Magyarkuti2018}. Note, that in the original studies \cite{Adak2015,Magyarkuti2018} a more sophisticated correlation method was used to determine the best fitting scaling exponent for the $S_I\sim G^n$ relation, but here we have replotted the data following the simplified $\Delta R/R$ vs. $R$ scheme also applied in the other panels. (c) In graphene nanogaps first a low level, resistance independent relative noise is observed due to the fluctuations in the leads (light red region), afterwards, rapidly increasing relative resistance fluctuations are detected due to atomic-scale junction width fluctuations (orange region together with the illustrative orange fitting line with a slope of unity on the $\log -\log$ scale), and finally, a saturated relative noise level is observed in the nanogap regime (yellow region). Note, that the $R_J$ junction resistance is obtained by subtracting the non-negligible resistance of the leads from the total $R$ device resistance. Data replotted from Ref. \citenum{Posa2021}. (d) The resistance dependence of the relative noise level in Ag$_2$S resistive switching junctions for diffusive filaments (light red region), ballistic filaments (orange region) and broken filaments (yellow region). The red and orange line-pair demonstrates the best fitting resistance scaling function according to the point-contact model describing the diffusive to ballistic crossover (see text). Data taken from Ref.~\citenum{Santa2019} completing the former results by datapoints in the yellow, broken filament regime. All data in the figure rely on noise measurements performed in the linear transport regime.}
  \label{Fig3}}
\end{figure}

In a nanoelectronic device, however, the functionality is confined to an extremely small volume where the above bulk considerations are clearly invalid. First of all, the relative contribution of a fluctuator heavily depends on its \emph{distance} from the device \emph{bottleneck}, i.e. a single or a few fluctuators positioned in the central active region may yield significantly larger noise contributions than the entire ensemble of more remote fluctuators. Furthermore, classical circuit rules loose their validity in devices entering the ballistic or quantum transport regimes.\cite{Elke_review} All these yield a non-obvious, highly system-specific scaling of the relative noise level with the device resistance. Confronting this scaling with possible transport models, one can identify the relevant transport mechanisms and noise sources in the studied system. The panels of Fig. \ref{Fig3} exemplify this scheme for different nanoelectric systems: (a) tunnel junctions between atomically sharp metallic apexes; (b) single-molecule nanowires (c) graphene nanojunctions and nanogaps; and (d) resistive switching memories. 

As the simplest reference system, the resistance of an \emph{atomic tunnel junction}, like a vacuum gap between an STM tip and a sample is an exponential function of the $d$ gap-size, $R\sim \alpha\cdot \exp(\beta\cdot d)$. A constant $\Delta d$ gap-size fluctuation, like subatomic fluctuations of the metallic apexes, yield a resistance fluctuation $\Delta R=\beta\cdot R\cdot \Delta d$. This yields a constant (resistance independent) relative resistance fluctuation, $\Delta R/R$, as confirmed by the measured resistance scaling in Fig. \ref{Fig3}a and in Ref.~\citenum{Adak2015}. It can be generally stated, that in systems, where the resistance is an exponential function of the physically fluctuating parameter, a mostly resistance independent relative resistance fluctuation is expected.

The noise characteristics of \emph{single-molecule nanowires},\cite{Tsutsui2010,Sydoruk2012,Xiang2014,Brunner2014,Karimi2016,Tewari2019,Kim2021}  where two atomically sharp metallic apexes are connected through a single organic molecule, can be analyzed by similar considerations. It was shown, that the resistance scaling of the noise is specific to the nature of the molecule-metal coupling, i.e. whether the electronic orbitals responsible for charge
transfer also participate in the formation of a chemical bond or
not.\cite{Adak2015} In the latter case (so-called through space coupling) the metal-molecule interface can show similar distance fluctuations as a tunneling junction, and therefore a more or less resistance independent relative resistance fluctuation is observed ($\Delta R/R\sim \textrm{const.}$). However, in the former case, the chemical bond ensures a rigid coupling, and therefore the noise is not related to distance fluctuations in the bond, rather to close-by atomic fluctuations of the metallic apex. This phenomenon results in a definite increase of the relative resistance fluctuation with the junction resistance yielding a scaling exponent close to $\Delta R/R\sim R^{0.5}$. Similar considerations could make difference between monomer single-molecule junctions with through-bond coupling on both sides, or a dimer junction, where the inter-molecular coupling has through-space nature.\cite{Magyarkuti2018} This is demonstrated in Fig. \ref{Fig3}b showing weakly resistance dependent relative noise in the high resistance, yellow region, where the dimer junctions yield through-space inter-molecular coupling, and a pronounced resistance dependence in the low resistance, light red region, corresponding to through-bond coupled monomer single-molecule junctions. Additionally, this distinction of the resistance scaling of the noise could detect molecular folding, where the folded molecule experienced a parallel through-space current noise between the gathering phenyl rings, which is absent in an unfolded junction.\cite{Wu2020} 

The noise characteristics also deliver fundamental information about two-dimensional devices, like \emph{graphene nanogaps}\cite{Puczkarski2018,Posa2021} or graphene single-electron transistors.\cite{Fried2020} The former system is exemplified in Fig. \ref{Fig3}c following our results in Ref.~\citenum{Posa2021}. A nanofabricated graphene wire (see left inset) is gradually narrowed by a feedback-controlled electrical breakdown procedure,\cite{Nef2014,Lau2014,ElAbbassi2017} obtaining atomic-sized graphene junctions (middle inset) and finally truly nanometer-wide nanogaps (right inset). Afterwards these nanogaps can be utilized to contact ultra-small functional elements, like single-molecules,\cite{ElAbbassi2019, QizhiXu2017, JanAMol2015, Lau2016} DNA sequences\cite{Heerema2016, Puczkarski2017} or ultra-small resistive switching filaments.\cite{Ashkan2015, Posa2017, Sarwat2017} In the first (light red) resistance regime the weakly resistance dependent and rather low noise level originates from the leads. However, as the junction resistance further increases, and the graphene junction narrows towards the ultimate atomic dimensions, a rapid, two orders of magnitude increase of the relative noise level is observed. In this regime the $R_J$ resistance of the junction is sensitive to the fine details of edge termination features, the precise position of nearby scattering centers and quantum interference phenomena, and so it is hardly described by simple theoretical considerations. However, combined conductance and transmission electron microscopy measurements\cite{Qi2014,Wang2016} delivered an inversely proportional empirical  dependence on the $W$ junction width: $R_J=\rho\cdot W^{-1}$. From this, a relative resistance fluctuation of $\Delta R/R=\Delta W \cdot R_J/\rho$ is obtained, yielding a slope of $1$ on the $\log(\Delta R/R)$ vs. $\log(R_j)$ plot, in full agreement with the experimentally observed noise increase in the orange region. The fitting of the noise data also delivers the $\Delta W$ amplitude of the width fluctuation. In this case $\Delta W\approx0.27\,$nm is obtained, i.e. truly atomic-sized junction width fluctuations, like atomic fluctuations at the junction edge are responsible for the observed noise characteristics. Entering the even higher (yellow) resistances regime, a clear crossover is observed in the noise characteristics, the rapidly increasing relative noise levels in the orange regime turn to a saturated relative noise in the yellow regime. Physically this crossover signals the transition from the unbroken nanojunction regime to the regime of broken nanogaps, where a tunneling current flows between the two sides. As noted above, in the tunneling regime the resistance is an exponential function of the possible fluctuating parameters, like the $d$ length of the tunneling gap or the $\Phi$ height of the tunneling barrier, and therefore, the saturated tendency of the relative noise levels is indeed expected.

The resistance scaling of the low-frequency noise also carries fundamental information in \emph{resistive switching memory devices, or memristors}.\cite{Ielmini2010,Soni2010,Fang2013,Ambrogio2014,Ambrogio2015,Yi2016,Puglisi2018,piros2020,Santa2019,Santa2021}  In these structure, the operation usually relies on the voltage-controlled formation and degradation of ultrasmall, truly nanometer-sized conducting filaments. Such a tiny active region is expected to be sensitive even to the noise generated by a single nearby atomic fluctuator. Indeed, many works reported telegraph noise in resistive switching filaments,\cite{Ielmini2010,Soni2010,Choi2014,Ambrogio2014,Ambrogio2015,Gonzalez2015,Gonzalez2016,Puglisi2018,Brivio2019} and the resistance scaling of the relative noise level was analyzed in terms of classical transport models relying on a metallic cylinder (or prism) geometry with a single two-state fluctuator at the filament surface perturbing the local effective cross-section.\cite{Ielmini2010,Ambrogio2014,Ambrogio2015,Puglisi2018} In this case, the relative resistance fluctuation scales with the relative cross-section fluctuation of the bottleneck, $\Delta R/R\sim \Delta A/A$,\cite{Ielmini2010} so a resistance scaling of $\Delta R/R\sim R$ is obtained according to the $R=\rho L/A$ relation, where $\rho$ is the resistivity, and L is the length of the cylindrical filament. 

Relying on past achievements in the field of point-contact spectroscopy,\cite{Jansen1980,Halbritter2004} it is also possible to describe a broader ensemble of fluctuators in the filament such that the suppression of a fluctuator's noise contribution by the increasing distance from the device bottleneck is explicitly taken into account considering a point-contact geometry. This model, however, is strongly dependent on the nature of electron transport in the filament.\cite{Wu2008,Santa2019} In ballistic devices, where the $d$ junction diameter is smaller than the $l$ mean free path, the junction conductance is described by the $G=G_0k_F^2d^2/16$ Sharvin formula,\cite{Sharvin} where $G_0=2e^2/h$ is the conductance quantum and $k_F$ is the Fermi wavenumber, whereas the contribution of a remote dynamical defect to the fluctuation of the conductance scales with the square of the $\Omega(\mathbf{r})$ solid angle, at which the junction is seen from the position of the fluctuator, $\Delta G_\mathrm{ballistic}\sim \Omega^2(\mathbf{r})$.\cite{Halbritter2004} In the diffusive regime ($l<d$) the junction conductance is determined by the $G=d/\rho$ Maxwell formula,\cite{Maxwell1891} and the diffusive nature of the transport yields an even stronger suppression of a remote fluctuator's contribution than in the ballistic regime, $\Delta G_\mathrm{diffusive}\sim \Omega^2(\mathbf{r})\cdot (l/d)^2$. Note, that these results rely on a point-contact geometry, where the filament diameter is gradually widening with the distance from the filament bottleneck, and therefore there is no characteristic filament length, the junction diameter is the single relevant length-scale in the system. The ballistic and diffusive transport regimes yield different scaling exponents between the relative noise level and the device resistance, which is indeed resolved in the measured resistance scaling. Fig.~\ref{Fig3}d demonstrates this diffusive-ballistic crossover of the noise in Ag$_2$S resistive switching junctions. In the light red (orange) diffusive (ballistic) regions respectively $(\Delta R/R)_\mathrm{diffusive}\sim R^{3/2}$ and $(\Delta R/R)_\textrm{ballistic}\sim R^{1/4}$ resistance scaling powers are observed in agreement with the model calculations.\cite{Santa2019} The fitting of the noise data also delivers an estimate on the $l$ mean free path and the $l_F$ average displacement of the fluctuators yielding nanometer scale values for both parameters.  Note, that the above diffusive resistance scaling power of $3/2$ can be simply motivated\cite{Wu2008} by considering the previously discussed $(\Delta R/R)^2\sim 1/V$ volume dependence in bulk samples, but identifying $V$ with the $\approx d^3$ central active volume, and using the $R\sim 1/d$ Maxwell relation. This simple argument, however, does not account for dependence on $l$ and $l_F$. Similar analysis of the noise's resistance scaling in Ta$_2$O$_5$ and Nb$_2$O$_5$ transition metal oxide resistive switching structures yields similar diffusive-ballistic crossover, but the overall noise levels are more than an order of magnitude smaller than in silver-based filaments.\cite{Santa2021} This material-specific noise reduction is attributed to the higher order of disorder, which yields an even stronger suppression of the remote fluctuators' contribution. 

At even higher resistance, i.e. above the $G_0^{-1}=h/(2e^2)\approx12.9\,$k$\Omega$ inverse conductance quantum,  broken filaments are envisioned, i.e. the metallic transport regime is replaced by electron tunneling, electron hopping, or temperature activated Poole-Frenkel phenomenon.\cite{Chiu2014} In these regimes, again the resistance is expected to be an exponential function of the possible  fluctuating parameters, and therefore, a saturated relative resistance fluctuation is expected. This saturated relative noise level in the non-metallic regime was indeed reported in various studies,\cite{Ielmini2010,Fang2013,Ambrogio2014,Ambrogio2015,Puglisi2018} and it is also demonstrated by the yellow region in Fig.~\ref{Fig3}d.

\section{Frequency scaling of the noise}
\label{freqscale}
In addition to the resistance scaling of the noise, the distinct frequency dependence of the noise power also carries fundamental information about the devices under study. According to Fig. \ref{Fig1}a a $1/f^2$ frequency scaling is expected in systems, where the investigated frequency range is above the inverse characteristic times of a single or more fluctuators generating the noise. As a clear distinction, a $1/f$-type frequency scaling (i.e. $1/f^\gamma$ with $\gamma\approx 1)$ rather signals the added contribution of multiple fluctuators with different time constants, such that the inverse time constants of the contributing fluctuators lie in the frequency range of the measurement. It is to be emphasized, that the added contribution of multiple fluctuators with different time constants indeed yields a significantly shallower frequency scaling compared to $1/f^2$, however, the actual value of the $\gamma$ scaling exponent depends on the distribution of the time constants in the actual device. As specific examples, the noise spectra of Ag$_2$S resistve switching memories (see Fig. \ref{Fig2}) exhibit an average scaling factor of $\gamma\approx1.12$, which is not far off from the \emph{ideal} $1/f$ scaling. The atomic point-contacts, tunnel junctions and single-molecule structures in Refs. \citenum{Adak2015} consequently exhibit a significantly larger scaling exponent of $\gamma\approx1.4$, which is attributed to configuration changes in the electrode apex due to
electrode atoms fluctuating between
metastable positions.  

The crossover between $1/f$-type and $1/f^2$-type frequency scaling was also resolved in various systems, for instance, in larger Ta$_2$O$_5$ resistive switching filaments a $1/f$-type scaling was observed, wheres for filaments approaching the single-atom diameter rather a $1/f^2$ scaling was reported.\cite{Yi2016}

Nanoelectronic devices with extremely small device volume may also exhibit a \emph{unique frequency scaling}, where a Lorentzian spectrum is superimposed on a $1/f$-type background,\cite{Santa2021,Puczkarski2018,Posa2021} as demonstrated by the bottom black spectrum in Fig.~\ref{Fig4}a. The decomposition of this mixed spectrum to a Lorentzian term and a $1/f^\gamma$ term permits to analyze how the noise of single  nearby fluctuators becomes dominant over a broader ensemble of more remote fluctuators as the active device volume is reduced.
This phenomenon was clearly observed in Nb$_2$O$_5$ and Ta$_2$O$_5$ transition metal oxide resistive switching filaments (the black spectrum in Fig.~\ref{Fig4}a. represents the latter material system),  where the contribution of remote fluctuators is suppressed by the enhanced degree of disorder, and therefore the Lorentzian spectrum of a single fluctuator positioned close to the filament bottleneck gives a significant contribution to the entire spectra.\cite{Santa2021} As a sharp contrast, in silver-based resistive switching filaments with larger mean free path, the disorder-induced noise suppression is less pronounced, yielding  significantly larger overall noise levels and the absence of a dominant Lorentzian contribution in the spectra (see the blue curve in Fig.~\ref{Fig4}a).\cite{Santa2021} 

Similar mixed spectra were also resolved in molecular electronic structures\cite{Kim2010,Hwang2013} and graphene nanogap devices, in the latter the Lorentzian contribution of single nearby fluctuators becomes significant either by approaching the ultimate atomic size-scales,\cite{Posa2021} or by decreasing the temperature.\cite{Puczkarski2017}   

\section{Nonlinear noise spectroscopy}
\label{nonlin}
So far, the linear, resistive transport regime was discussed, i.e. devices with linear current-voltage characteristics, $I=V/R$. Furthermore, steady state resistance fluctuations were considered as the source of the $1/f$-type noise yielding a quadratic dependence of the $S_I$ noise power on the driving voltage amplitude (see Fig.~\ref{Fig2}d), which yields constant, voltage independent relative relative current, voltage or resistance fluctuations. The deviation of the noise power from the squared dependence, however, carries additional information on the devices under study.  

\begin{figure}[!h]
  \centering{\includegraphics[width=0.48\textwidth]{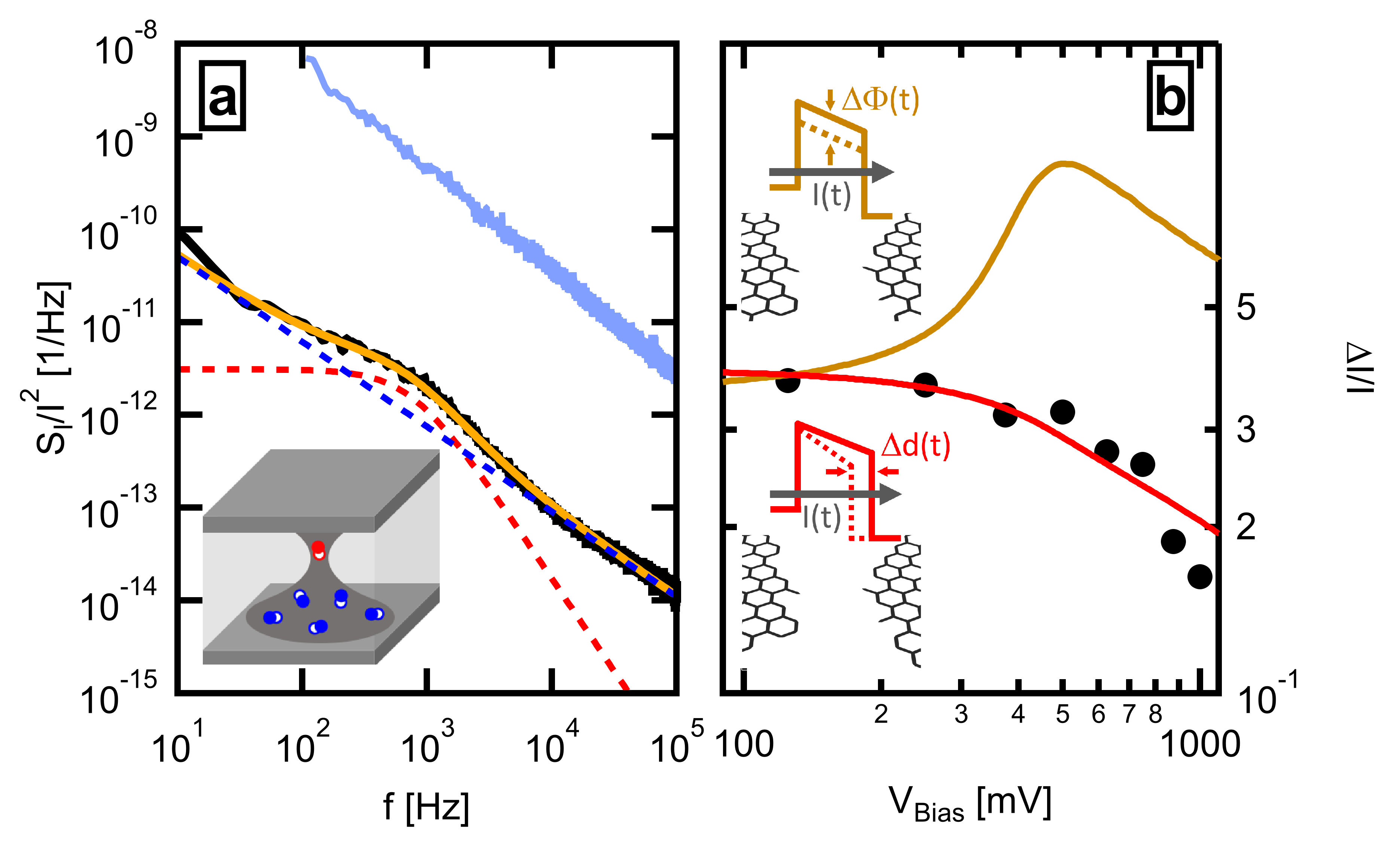}
  \caption{{\bf Frequency scaling of the noise and nonlinear noise spectroscopy.} (a) Representative spectra demonstrating a clear $1/f^\gamma$ frequency scaling (blue curve), and a mixed spectrum (black curve), which is well fitted (yellow line) by the added contribution of a $1/f^\gamma$ spectrum of an ensemble of remote fluctuators (blue dashed line) and a Lorentzian spectrum of a single nearby fluctuator (red dashed line). The inset illustrates a (red) nearby fluctuator at the filament bottleneck and an ensemble of more remote fluctuators (blue). The black curve is measured in the LRS ($R=295\,\Omega$) of a Ta$_2$O$_5$ resistive switching junction, similar to the measurements in Ref.~\citenum{Santa2021}. The blue spectrum is measured on the LRS ($R=300\,\Omega$) of a Ag$_2$S resistive switching junction (the same spectrum as the $V_\mathrm{bias}=27\,$mV spectrum in Fig.~\ref{Fig2}b). Thanks to the similarity of the two resistances, it is clear from the offset of the two spectra that the Ag$_2$S system exhibits orders of magnitude larger noise level than the Ta$_2$O$_5$ system. (b) The voltage dependence of the relative current fluctuation in the nonlinear transport regime of a graphene nanogap device (black dots).\cite{Posa2021} The red and light brown curves demonstrate the expected voltage dependent relative current fluctuations in case of gap-size fluctuations and barrier height fluctuations, respectively (see the illustrative insets for the two models).}
  \label{Fig4}}
\end{figure}

Even in a linear resistive system, the voltage driving may introduce further fluctuations in addition to the steady state resistance noise. In this case, the $S_I\sim V^2$ relation is violated, or in other words, the relative current fluctuation, $\Delta I/I$ is not constant, but it is increasing by the voltage. As an example, the voltage induced fluctuations were clearly resolved in low-temperature atomic tunnel juctions due to the electrostatic lowering of the potential barrier for the two-level fluctuations.\cite{Adak2015} Such nonlinear noise phenomenon can be applied as a spectroscopic tool to resolve voltage induced fluctuations in various nanoelectric devices. Note, that the detection of enhanced, voltage-induced fluctuations in the active volume
may also precursor forthcoming drastic changes in the conductance, like resistive switching, local electric-field induced phase change, etc.

As a second option, one can consider a nanoelectronic device with definite current-voltage nonlinearity. The nonlinear $I\left(V\right)$ curve of the system may be converted to a nontrivial driving dependence of the noise. As an example, the nonlinear transport in a graphene tunnel junction is described by the Simmons model,\cite{Simmons1963} where the $I\left(V\right)$ curve depends both on the $\Phi$ work function and on the $d$ nanogap-size. However, this nonlinear current converts to nonlinear noise in a different fashion if the work function is fluctuating due to charge traps, or if the gap-size is fluctuating due to mobile adatoms, i.e. noise measurements may distinguish between different physical processes. This scheme is exemplified in Fig.~\ref{Fig4}b, where the black dots represent the measured relative current fluctuation values in the nonlinear regime, whereas the red and light brown lines respectively demonstrate the expected voltage dependence of the relative noise level in case of barrier-height or gap-size fluctuations (see the illustrative insets). It is clear, that the voltage dependence of the former model is fully inconsistent with the measured data, whereas the gap-size fluctuation model with subatomic $\Delta d\approx0.05\,$nm describes the data well.\cite{Posa2021}

Similar concept was applied to analyze the nonlinear noise spectra in phase-change type resistive switching memory devices, where the the transport nonlinearity is described by the Poole-Frenkel mechanism.\cite{Beneventi2009} 

\section{Noise tailoring}
\label{tailor}

The commercial introduction of a novel electronic device is often preceded by a lengthy material optimization phase devoted to the suppression of the low-frequency device noise as much as possible. The above summarized noise anyalysis concepts may serve helpful for this purpose through the proper identification of the dominant noise sources in the actual device. The emergence of novel computing architectures, however, triggers a paradigm change in noise engineering, demonstrating that a non-suppressed, but tailored noise can be harvested as a computational resource in probabilistic computing schemes. 

\begin{figure}[!h]
  \centering{\includegraphics[width=0.48\textwidth]{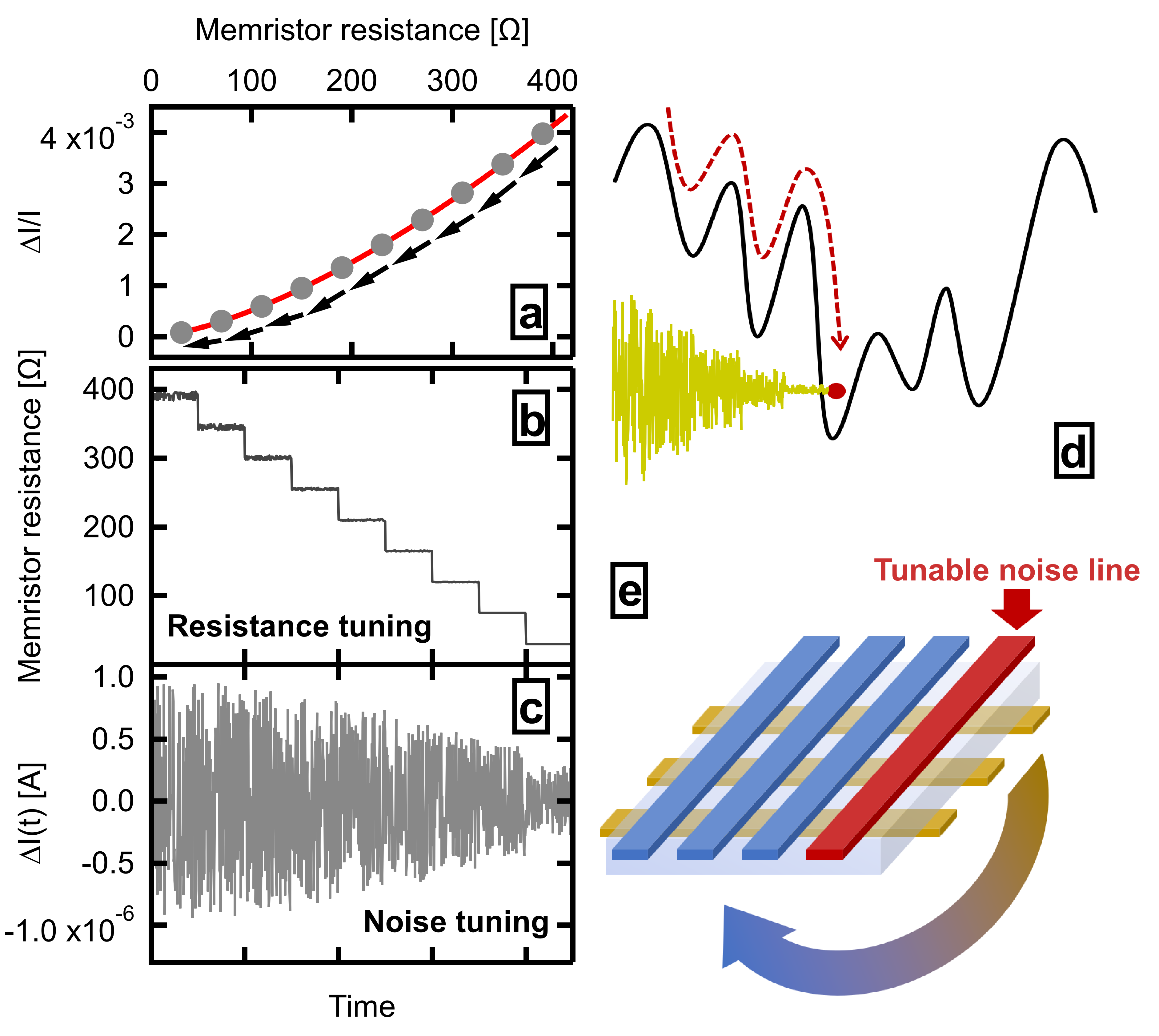}
  \caption{{\bf Noise tailoring.}  
The fictive data in panels (a,b,c) illustrate the envisioned scheme of noise tuning (see text). Panel (a) illustrates the resistance scaling of the noise for diffusive resistive switching filaments (note the linear scale on both axis), and the gradual decrease of the device resistance by proper voltage manipulation (see the grey dots and the black arrows) . Panels (b) and (c) illustrate the gradual resistance decrease, and the gradual current noise decrease along this process. Note, that these panels illustrate realistic resistance values in the diffusive regime of Ag$_2$S resistive switching junctions (see Fig.~\ref{Fig3}d), and related realistic current noise levels considering a constant $V=100\,$mV bias voltage on the junction. Panel (d) illustrates the benefit of noise tuning in a probabilistic computing scheme: the initially larger noise levels help to escape from local minima in the energy landscape of the problem, whereas the gradual decrease of the noise level aids the convergence to the global, energy minimized solution. (e) According to the proposal in Ref.~\citenum{cai2019harnessing} a Hopfield neural network may be extended by an additional line with tunable noise characteristics. The base crossbar executes the key vector-matrix multiplication operation of the neural network\cite{Xia2019} (note, that according to Ohm's law and Kirchhoff's rule, the output current vector at the horizontal light brown lines is simply the product of the input voltage vector on the blue lines, and the conductance matrix of the memristors positioned at the crosspoints). Such a memristor crossbar -- calculating the vector-matrix product in a single time-step -- serves as a hardware accelerator compared to the computation intensive software approaches. In a Hopfield network a proper neural operation feedbacks the output to the input such that each iteration decreases the energy function of the problem encoded in the memristor weights. Instead of using an external circuitry for noise tuning,\cite{Cai2020} an additional tunable noise line (red) could be applied to implement the gradual noise decrease.\cite{cai2019harnessing}}
  \label{Fig5}}
\end{figure}

As two demonstrative examples, networks of magnetic tunnel junctions\cite{Borders2019} as well as resistive switching memories\cite{Cai2020} were succesfully applied to solve nondeterministic polynomial-time (NP) hard computational problems on the hardware level. As a common feature, both works relied on noise tuning. 

In the former case the magnetic tunnel junctions served as probabilistic bits (p-bits), where the relative occupation of the two states was voltage-tunable in the random telegraph noise of the device. The proof of concept experiment with an eight p-bit circuit demonstrated prime number factorization.\cite{Borders2019}

In the latter case a 60x60 crossbar structure of resistive switching memory elements was applied to realize a Hopfied neural network, which finds the minima in the energy landscape of the targeted problem along its operation. An external circuitry was used to amplify/suppress the intrinsic noise of the network. The gradual decrease of the noise level (like a simulated annealing protocol) was applied to escape from local minima, and to find the global solution of the problem (see the illustration in Fig.\ref{Fig5}d). This scheme was used to solve max-cut problems of graphs claiming to deliver over four orders of magnitude higher solution throughput per power consumption than digital or quantum annealing approaches.\cite{Cai2020}

These examples demonstrate a novel perspective for noise engineering, where not the suppression of the noise is the key target, rather the material optimization should deliver noise characteristics, which are best suited for the targeted computational problem, and which are well tunable along the device operation. Resistive switching memories with high endurance and data retention are especially promising candidates for this purpose, as their resistance state is analog tunable with high precision,\cite{Li2017} and thereby their noise level is also tuned. This scheme is illustrated in  Figs.~\ref{Fig5}a,b,c. Panel (a) illustrates the resistance scaling tendency of the noise in diffusive metallic filaments.  By proper voltage pulses one can gradually decrease the device resistance (see the grey dots in Fig.~\ref{Fig5}a, and the fictive temporal evolution of the resistance in Fig.~\ref{Fig5}b), and thereby, the amplitude of the $\Delta I(t)$ current fluctuation at constant voltage driving also gradually decreases (Fig.~\ref{Fig5}c). According to the proposal in Ref.~\citenum{cai2019harnessing} such noise tuning may be implemented in a memristor crossbar structure by an additional tunable noise line, which adds tunable current noise to the Hopfield neural network (Fig.~\ref{Fig5}e). 

\section{Conclusions}

In conclusion, we have illustrated the merits of $1/f$-type noise analysis in nanoelectronic devices. We demonstrated, that the specific resistance scaling of the noise may differentiate between fundamentally different transport mechanisms, like through-space or through-bond coupled molecular transport, ballistic or diffusive metallic transport, or the transition from the nanojunction to the nanogap regime. Furthermore, the fitting of the noise's resistance scaling delivers relevant microscopic parameters, like the amplitude of junction width or gap-size fluctuations, the density of fluctuators, or the mean free path.  In the frequency scaling of the noise, the increasing dominance of single, nearby fluctuators over a broader ensemble of more remote fluctuators can be followed as the active volume of the device is decreased. As a third approach, nonlinear noise spectroscopy serves as a useful tool to recognize voltage-induced fluctuations over the steady-state resistance fluctuations, and additionally, the combined analysis of the current-voltage nonlinearity and the non-obvious voltage dependence of the noise helps to identify the relevant fluctuating parameter in the transport model. Finally, we have briefly reviewed the recent progress in the development of probabilistic computing hardware machines, where the well-tunable noise of the devices is a key ingredient of the operation.  

\section*{Acknowledgements}

We are thankful to L. Venkataraman for sharing the noise data on the dimer and monomer molecular junctions and for useful discussions on the single-molecule noise measurements. This work was supported by the Ministry of Innovation and Technology and the National Research, Development and Innovation Office within the Quantum Information National Laboratory of Hungary, and the NKFI K128534 and K119797  grants. 

\section*{Author contributions}

The manuscript was written by A. Halbritter and Z. Balogh. G. Mezei, A. Magyarkuti, L. Pósa and B. Sánta respectively contributed to the measurement, analysis and illustration of the noise data on atomic tunnel junctions, dimer and monomer molecular junctions, graphene nanogaps and resistive switching memory junctions.  

\balance

\bibliography{noise_review_refs}
\bibliographystyle{rsc}
\end{document}